\documentclass[11pt]{amsart}
\usepackage{graphicx}
\usepackage{hyperref}
\theoremstyle{definition}

\theoremstyle{remark}

\numberwithin{equation}{section}
\def \dd{{\rm d}}
\begin{document}
\title{The issue of photons in
dielectrics: Hamiltonian viewpoint}
\author{S. Antoci and L. Mihich}
\address{Dipartimento di Fisica ``A. Volta'' and IPCF of CNR,
Via Bassi 6, Pavia, Italy}
\email{Antoci@fisicavolta.unipv.it}

%\thanks{}%
%\keywords{General relativity, electrodynamics, geometrical optics, light quanta}%

\begin{abstract}
The definition of the photon in the vacuum of general relativity
provided by Kermack et al. and by Synge is extended to
nondispersive, nonhomogeneous, isotropic dielectrics
in arbitrary motion by Hamiltonian methods that rely
on Gordon's effective metric.
By these methods the old dilemma, whether the momentum-energy
vector of the photon in dielectrics is timelike or spacelike in character,
is shown to reappear under a novel guise.
\end{abstract}
\maketitle
% ----------------------------------------------------------------
\section{Introduction}
Despite a widespread conviction, the concept of photon in vacuo
does not pertain exclusively to quantum physics. Classical general
relativity has its say on the subject, as it was beautifully shown
long ago by Kermack et al. \cite{Whittaker1933} and by
Synge \cite{Synge1935}. By availing of Hamiltonian methods, and of
the reduction to the vacuum case of the geometrical optics of
dielectrics operated by Gordon \cite{Gordon1923} and by Pham Mau
Quan \cite{Pham Mau Quan1957}, the problem of the
definition of the photon in dielectrics is considered
anew in the present paper.

The abstract formulation \cite{Synge1960} of Hamilton's theory of
rays and waves is first recalled (Section 2). The metric is then introduced,
and the particular Hamiltonians that apply to the case of the
vacuum (Section 3), and of isotropic, nondispersive dielectrics
(Section 4) are considered. Synge's derivation of the
proportionality between energy and frequency for the photon in the
vacuum of general relativity is then displayed (Section 5). His
argument, when applied to the case of dielectrics, leads to
a twofold possibility: one must choose either a timelike photon
(Section 6), for which however the proportionality between energy and frequency
breaks down in nonhomogeneous dielectrics, or a spacelike
photon (Section 7) for which the proportionality is always
ensured. The latter option however requires, at variance with what occurs
in mechanics, that an entity endowed with a spacelike momentum-energy
vector be associated to a timelike ray. The existence of this
twofold possibility is akin to the ancient, unsolved dilemma in the
electromagnetism of continua, whether the energy tensor of the
electromagnetic field in dielectrics should be given either the
expression proposed by Abraham \cite{Abraham1909} or the one
formulated by Minkowski \cite{Minkowski1908}.
\section{Hamilton's theory in abstract form}
Like Maxwell's equations, also the equations of Hamilton's theory
of rays and waves \cite{Synge1960} lie very deep in the conceptual
structure of physics. In fact, as it occurs with Maxwell's equations
\cite{Schroedinger60}, these equations can be written without availing
of either an affine or a metric connection: a bare eight-dimensional
manifold suffices, equipped only with coordinates $x^i$ and $y_i$
\footnote{Latin indices run henceforth from 1 to 4.}. In this
manifold a 7-surface $\Sigma$ is assumed to exist, whose equation
is
\begin{equation}\label{2.1}
H(x^i,y_i)=0.
\end{equation}
By following Synge's account, we consider a curve $\Gamma$ joining
the points $A$ and $B$ on $\Sigma$, and the integral
\begin{equation}\label{2.2}
I=\int_A^By_i\dd x^i.
\end{equation}
We propose ourselves to find the extremals of this integral with
the varied curve always lying on $\Sigma$. To this end a parameter $u$
is introduced, whose values at $A$ and at $B$ are fixed
for all the curves, and we get rid of the side condition
(\ref{2.1}) by considering the variation of the integral
\begin{equation}\label{2.3}
J=\int_A^B(y_i\dd x^i-\lambda H\dd u),
\end{equation}
where $\lambda(u)$ is a Lagrange multiplier. A generic variation
of $J$ reads
\begin{eqnarray}\label{2.4}
\delta J=\left[y_i\delta x^i\right]^A_B\\\nonumber
+\int_A^B\left(\delta y_i\dd x^i-\delta x^i\dd y_i
-H\delta\lambda\dd u
-\lambda\frac{\partial H}{\partial x^i}\delta x^i \dd u
-\lambda\frac{\partial H}{\partial y_i}\delta y_i \dd u\right).
\end{eqnarray}
It is asked that $\delta J=0$ under arbitrary variations of
$\delta x^i$, $\delta y_i$, $\delta\lambda$, provided that
$\delta x^i$ vanish at $A$ and at $B$. One finds that the extremals
must obey the equations
\begin{equation}\label{2.5}
\frac{\dd x^i}{\dd u}=\lambda\frac{\partial H}{\partial y_i},\;
\frac{\dd y_i}{\dd u}=-\lambda\frac{\partial H}{\partial x^i},\;
H=0.
\end{equation}
Since the variation does not determine $\lambda(u)$, one can
choose the parameter $u$ in order to write the equations of the
extremals in the Hamiltonian form
\begin{equation}\label{2.6}
\frac{\dd x^i}{\dd u}=\frac{\partial H}{\partial y_i},\;
\frac{\dd y_i}{\dd u}=-\frac{\partial H}{\partial x^i},\;
H=0.
\end{equation}
These equations can be given another interpretation, in terms of a
four-dimensional manifold whose coordinates are $x^i$. In the new
interpretation, equation (\ref{2.1}) no longer refers to a
7-surface $\Sigma$, but to a set of 3-spaces, each one associated
to a point $x^i$. We want that the theory be invariant under
arbitrary transformations of the $x$-coordinates, hence we must
think of the $y_i$ as a covariant four-vector defined on the
manifold. Considered in the original eight-dimensional space, an
extremal is a curve on $\Sigma$. In the four-dimensional manifold
whose coordinates are $x^i$, it appears as a curve $x^i=x^i(u)$
with an associated vector field $y_i=y_i(u)$. Let us consider from
the four-dimensional standpoint the integral (\ref{2.2}) along an
extremal joining the points $A({x'}^i)$ and $B(x^i)$, and call it
$f({x'}^i, x^i)$. It is nothing but Hamilton's characteristic
function. If the points $A$ and $B$ are varied and the integral
for the new extremal is compared to the integral for the previous
one, since the integral in the second row of (\ref{2.4}) vanishes
for the extremals, one finds
\begin{equation}\label{2.7}
\delta f=\left[y_i\delta x^i\right]^B_A.
\end{equation}
Provided that $\delta{x'}^i, \delta{x}^i$ can be chosen
arbitrarily, it must be
\begin{equation}\label{2.8}
\frac{\partial f}{\partial x^i}=y_i,\;
\frac{\partial f}{\partial {x'}^i}=-{y'}_i.
\end{equation}
Inserting the first of (\ref{2.8}) in (\ref{2.1}) produces the
Hamilton-Jacobi equation
\begin{equation}\label{2.9}
H\left(x^i,f_{,i}\right)=0,
~\text{where}~f_{,i}\equiv\frac{\partial f}{\partial x^i}.
\end{equation}
If the vector field $y_i$ defined by the extremals is such that
the circulation
\begin{equation}\label{2.10}
\oint_{C}f_{,i}dx^i=0
\end{equation}
for an arbitrary closed curve $C$ lying within some (two,
three or four dimensional) domain $D$, the extremals form what
Synge calls a coherent system and are named rays. For a coherent
system, the integral
\begin{equation}\label{2.11}
I(A,B)=\int_A^Bf_{,i}dx^i
\end{equation}
between a fixed point $A$ and a variable point $B$ does not depend
on the path of integration. The subspace of $D$ that is found when
$B$ varies but $I(A,B)$ maintains a constant value $v$ is called a
wave; by varying the parameter $v$ we get a set of waves; changing
the starting point $A$ does not change the set; it merely
changes its parametrisation.
\section{Introducing the metric. Null waves and null rays.}
For associating the quantities of the theory with the quantities
of experience we need a metric. We assume that $g_{ik}$ is the
usual pseudo-Riemannian metric of general relativity, that can be
locally reduced through a suitable transformation to the Minkowski
form $\eta_{ik}\equiv \text{diag}(1,1,1,-1)$. Then we can
introduce the timelike world-line of an observer endowed with
four-velocity $u^i$ and define the frequency of the wave with
respect to that observer:
\begin{equation}\label{3.1}
\nu=f_{,i}u^i.
\end{equation}
We can also consider whether $f_{,i}$ is spacelike, timelike or
null. In the first case $\nu$ can vanish
locally, while this is impossible in the second case. One can also
show \cite{Synge1960} that in the first case the phase velocity of the wave is
inferior to the fundamental velocity introduced by the metric,
while the converse is true in the second case. In the third case
the wave propagates just with the fundamental velocity. Such a null
wave obeys the Hamilton-Jacobi equation
\begin{equation}\label{3.2}
H\left(x^i,f_{,i}\right)=\frac{1}{2}g^{ik}f_{,i}f_{,k}=0.
\end{equation}
Since in this case the first of Hamilton's equations (\ref{2.6}) reads
\begin{equation}\label{3.3}
\frac{\dd x^i}{\dd u}=g^{ik}f_{,k}
\end{equation}
the ray velocity $\frac{\dd x^i}{\dd u}$ is null:
\begin{equation}\label{3.4}
g_{ik}\frac{\dd x^i}{\dd u}\frac{\dd x^k}{\dd u}=0.
\end{equation}
The second of Hamilton's equations reads now
\begin{equation}\label{3.5}
\frac{\dd f_{,i}}{\dd u}=-\frac{1}{2}g^{kl}_{~,i}f_{,k}f_{,l}.
\end{equation}
By eliminating $f_{,i}$ in (\ref{3.5}) with the use of (\ref{3.3})
one finds that
\begin{equation}\label{3.6}
\frac{\delta}{\delta u}\frac{\dd x^i}{\dd u}=0;
\end{equation}
the extremals for the Hamiltonian (\ref{3.2}) are null
geodesics, with the parameter $u$ of the extremal taking the role
of special parameter for the geodesic.\par
For a coherent system one has
\begin{equation}\label{3.7}
f_{,i}\delta x^i=0
\end{equation}
when $\delta x^i$ is a displacement along the wave. Due to (\ref{3.3})
one obtains
\begin{equation}\label{3.8}
g_{ik}\delta x^i\frac{\dd x^k}{\dd u}=0.
\end{equation}
Since, again due to (\ref{3.3})
\begin{equation}\label{3.9}
f_{,i}\frac{\dd x^i}{\dd u}=g^{ik}f_{,i}f_{,k}=0
\end{equation}
one finds that not only the null ray is orthogonal to the null
surface, but it lies in it too. The fact that a certain curve
lies in a certain surface is however an occurrence that can be
ascertained without a metric; we shall remind of this property in
the following, when dealing with photons in dielectrics.
\section{The characteristics of Maxwell's equations for a
nondispersive, isotropic dielectric}
We define \cite{Post1962} the electric displacement and
the magnetic field by the antisymmetric, contravariant tensor
density ${\bf H}^{ik}$, while the electric field and the magnetic
induction are given by the skew, covariant tensor
$F_{ik}$. We define also the four-vectors:
\begin{equation}\label{4.1}
F_{i}=F_{ik}u^{k},~~~H_{i}=H_{ik}u^{k},
\end{equation}
where $u^i$ is the four-velocity of the medium. If the latter is isotropic
in its rest frame, its linear constitutive equation reads \cite{Minkowski1908}
\begin{equation}\label{4.2}
\mu{H^{ik}}=F^{ik}+(\epsilon\mu-1)(u^{i}F^{k}-u^{k}F^{i}).
\end{equation}
Gordon \cite{Gordon1923} noticed that (\ref{4.2}) can be rewritten as
\begin{equation}\label{4.3}
\mu{H^{ik}}=\left[g^{ir}-(\epsilon\mu-1)u^{i}u^{r}\right]
\left[g^{ks}-(\epsilon\mu-1)u^{k}u^{s}\right]F_{rs}.
\end{equation}
Therefore he introduced the ``effective metric tensor''
\begin{equation}\label{4.4}
\sigma^{ik}=g^{ik}-(\epsilon\mu-1)u^{i}u^{k}
\end{equation}
and rewrote the constitutive equation (\ref{4.2}) as
\begin{equation}\label{4.5}
\mu{\bf H}^{ik}=\sqrt{g}\sigma^{ir}\sigma^{ks}F_{rs},
\end{equation}
where $g\equiv-\det(g_{ik})$. Since after some simple algebra one finds
\cite{Gordon1923} that
\begin{equation}\label{4.6}
\sigma=\frac{g}{\epsilon\mu},
\end{equation}
where $\sigma\equiv-\det(\sigma_{ik})$, equation (\ref{4.3}) can be
eventually rewritten as
\begin{equation}\label{4.7}
{\bf H}^{ik}=\sqrt{\frac{\epsilon}{\mu}}
\sqrt{\sigma}\sigma^{ir}\sigma^{ks}F_{rs}.
\end{equation}
Therefore, apart from the factor $\sqrt{\epsilon/\mu}$,
the constitutive equation in a nondispersive, isotropic dielectric
medium is rendered by the effective metric $\sigma_{ik}$ exactly
in the way the constitutive equation for vacuum is rendered by the
pseudo-Riemannian metric $g_{ik}$. If the medium is also homogeneous
in its rest frame, the factor $\sqrt{\epsilon/\mu}$
is constant, and the analogy with the vacuum case is complete. Then one
can reduce to the vacuum the problem of the choice of the Lagrangian
for the electromagnetic field in nondispersive, homogeneous,
isotropic dielectrics. As a consequence, in
such dielectrics the definition of the energy tensor of the
electromagnetic field is uniquely given by Gordon's
procedure \cite{Gordon1923}, and in this way the general relativistic
extension of Abraham's form \cite{Abraham1909} is retrieved.
In his seminal paper Gordon, after
finding the expression of the wave equation for tensorial fields
in a pseudo-Riemannian space, dealt with the limit of geometrical
optics in the footsteps of Hadamard \cite{Hadamard1902} and proved
that the propagation of light in a dielectric that is isotropic
and homogeneous in its local rest frame must obey, besides an
equation that links the amplitude $A$ and the phase $f$, the equation
\begin{equation}\label{4.8}
\sigma^{ik}f_{,i}f_{,k}=0,
\end{equation}
where $\sigma^{ik}$ is the effective metric (\ref{4.4}). To this equation
Gordon applied the Hamiltonian method of Section 3 to prove
\footnote{See Ref. \cite{Gordon1923}, page 455.}, inter alia, that
``the rays in moving bodies are represented through null geodesics
of the manifold with the line element
$d\sigma^2=\sigma_{ik}dx^idx^k$''.

Gordon's proof applies to homogeneous dielectrics. In 1957, by
studying the characteristics of the Maxwell-Einstein equations
\cite{Pham Mau Quan1957}, Pham Mau Quan extended the validity of
(\ref{4.8}) to a nonhomogeneous medium.
\section{Photons in the vacuum of general relativity}
In two remarkable papers \cite{Whittaker1933,Synge1935} Kermack et
al. and Synge have shown that the concept of photon in vacuo is by
no means an exclusive outcome of quantum theory, but is deeply
rooted in classical general relativity. We have already availed
\cite{Antoci2001} of Synge's proof of the proportionality of energy and frequency for
an entity endowed with both the behaviour of a null wave and of a
null particle, exchanged between two atoms
in arbitrary states of motion in a gravitational field. We
have rather reproducing it here, since it is necessary for
the discussion of photons in dielectrics from the Hamiltonian viewpoint.\par
Atoms are dealt with as pointlike entities. In Figure 1 are shown the
world-lines of an emitting and of an absorbing atom, joined by a null line
$A_0A$. We attribute
to the photon the Hamiltonian (\ref{3.2}); viewed as a point particle, its equations
of motion are (\ref{3.6}) and (\ref{3.4}).

\begin{figure}[h]
\includegraphics{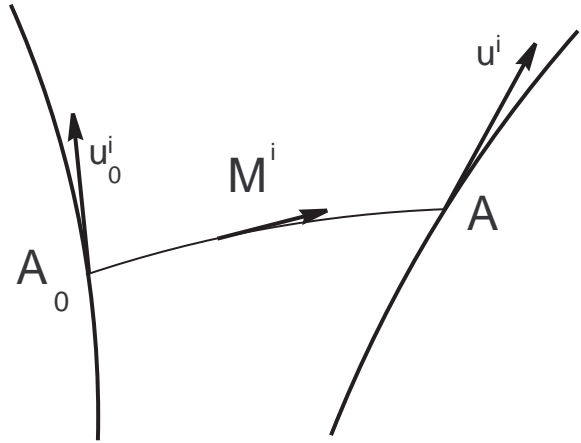}
\caption{Spacetime diagram for two atoms exchanging Synge's photon.}
\end{figure}
\noindent We endow this point particle with a momentum-energy vector $M^i$
that is tangent to the photon's world-line and is parallelly
propagated along it:
\begin{equation}\label{5.1}
M^i=\theta\frac{\dd x^i}{\dd u},\text{~~~}\frac{\delta M^i}{\delta
u}=0,
\end{equation}
where $\theta$, assumed to be positive, happens to be a constant
thanks to (\ref{3.6}). Since it travels with the fundamental
velocity, the photon must have zero proper mass, hence:
\begin{equation}\label{5.2}
M_iM^i=0.
\end{equation}
An atom is instead pictured as a timelike point particle endowed with
the momentum-energy vector
\begin{equation}\label{5.3}
m_0 u^i=m_0\frac{\dd x^i}{\dd s};
\end{equation}
$m_0$ is the atom's rest mass or rest energy, while $u^i$ is its
four-velocity. The change of the momentum-energy vector of the
atom due to absorption of a photon is supposedly ruled by the
conservation law:
\begin{equation}\label{5.4}
m'_0 {u'}^i=m_0 u^i+M^i,
\end{equation}
where $m'_0$ is the rest mass, and ${u'}^i$  is the four-velocity
of the atom after the absorption of the photon. By rewriting the
conservation law in covariant form and by multiplying the two
forms term by term one finds
\begin{equation}\label{5.5}
-{m'}_0^2=-m_0^2+2m_0M_iu^i,
\end{equation}
{\it i.e.}
\begin{equation}\label{5.6}
\frac{{m'}_0^2-m_0^2}{2m_0}=-M_iu^i.
\end{equation}
Since for optical processes $|{m'}_0-m_0|/{m_0}\ll 1$ Synge
defines
the energy $E$ of a photon with momentum-energy vector $M^i$
relative to an atom endowed with four-velocity $u^i$ as
\begin{equation}\label{5.7}
E=-M_iu^i.
\end{equation}
This scalar quantity is numerically equal to the fourth component
of $M^i$ in a locally Minkowskian coordinate system where the atom
is at rest.

The special parameter $u$ along the null geodesics joining the world-lines of the
absorbing and of the emitting atoms can be given a single starting value
and a single terminal value on these lines, since the geodesic equation
is invariant under the linear transformation $u'=au+b$, where $a$ and
$b$ are constants. Under this transformation the Hamilton
equations (\ref{3.3}) and (\ref{3.5}) are preserved too, provided that
the $y_i$ undergo the corresponding transformation. The latter does not
affect the structure of the waves, but only their parametrisation.
Let $v$ be the parameter introduced at the end of Section 2,
associated to the waves after this
reparametrisation has occurred. We can write
\begin{equation}\label{5.8}
\frac{\partial}{\partial u}
\left(M_i\frac{\partial x^i}{\partial v}\right)
=\frac{\delta M_i}{\delta u}\frac{\partial x^i}{\partial v}
+M_i\frac{\delta}{\delta u}\frac{\partial x^i}{\partial v}.
\end{equation}
We have also
\begin{equation}\label{5.9}
\frac{\delta}{\delta u}\frac{\partial x^i}{\partial v}
\equiv \frac{{\partial}^2 x^i}{\partial u\partial v}
+\Gamma^i_{kl}\frac{\partial x^k}{\partial v}
\frac{\partial x^l}{\partial u}
=\frac{\delta}{\delta v}\frac{\partial x^i}{\partial u},
\end{equation}
where $\Gamma^i_{kl}$ is the Christoffel connection built with
$g_{ik}$. Since $\delta M_i/\delta u=0$, the right hand side of
(\ref{5.8}) can be rewritten as
\begin{equation}\label{5.10}
M_i\frac{\delta}{\delta u}\frac{\partial x^i}{\partial v} =\theta
g_{ik}\frac{\partial x^k}{\partial u} \frac{\delta}{\delta
v}\frac{\partial x^i}{\partial u}
=\frac{1}{2}\theta\frac{\partial}{\partial v}
\left(g_{ik}\frac{\partial x^i}{\partial u} \frac{\partial
x^k}{\partial u}\right)=0.
\end{equation}
Hence one finds
\begin{equation}\label{5.11}
\frac{\partial}{\partial u}
\left(M_i\frac{\partial x^i}{\partial v}\right)=0,
\end{equation}
and, with reference to Figure 1:
\begin{equation}\label{5.12}
\left(M_i\frac{\partial x^i}{\partial v}\right)_{A_0}
=\left(M_i\frac{\partial x^i}{\partial v}\right)_A.
\end{equation}
Let $u^i_0$, $u^i$ be the four-velocities of the atoms at $A_0$
and at $A$ respectively. If $dv$ is the infinitesimal increment of
the parameter $v$ when going from the null surface joining $A_0$ and
$A$ to a neighbouring one, we can write:
\begin{equation}\label{5.13}
\left(\frac{\partial x^i}{\partial v}\right)_{A_0}dv
=u^i_0ds_0,\text{~~~}
\left(\frac{\partial x^i}{\partial v}\right)_Adv
=u^ids,
\end{equation}
hence from (\ref{5.12}) we get
\begin{equation}\label{5.14}
\left(M_iu^i\right)_{A_0}ds_0=\left(M_iu^i\right)_Ads,
\end{equation}
and the definition (\ref{5.7}) of the energy of a photon absorbed
(or emitted) by an atom allows rewriting the last equation as
\begin{equation}\label{5.15}
E_0ds_0=Eds.
\end{equation}
From the wave model of the photon implicit in Hamilton's
formulation in a coherent system one gathers that,
if $ds_0$ and $ds$ are the intervals
during which the emission and the absorption of the wavelike
photon takes place respectively at the atoms $A_0$ and $A$, it
must be:
\begin{equation}\label{5.16}
\nu_0ds_0=\nu ds,
\end{equation}
where $\nu_0$ and $\nu$ are the scalar frequencies
\footnote{The value of the scalar frequency (\ref{3.1}) depends on the
parametrisation, but this fact affects both sides of equation (\ref{5.16})
in the same way.} of emission and of
absorption defined by (\ref{3.1}).
By dividing term by term this equation and
equation (\ref{5.15}) we get:
\begin{equation}\label{5.17}
\frac{E_0}{\nu_0}=\frac{E}{\nu}.
\end{equation}
Therefore, Hamilton's theory allows to show that if a photon is emitted by
an atom and absorbed by another one in presence of a
gravitational field, the ratio energy/frequency is the same for
emission and for absorption. This ratio is independent of the
behaviour of the gravitational field and of the state of motion of
the two atoms.
\section{Timelike photons in nondispersive, isotropic dielectrics}
Thanks to Gordon's effective metric (\ref{4.4}) and to the
corresponding Hamiltonian (\ref{4.8}) it is immediate to extend
Synge's proof to nondispersive, iso\-tropic dielectrics. The
diagram in Figure 1 applies also in this case, and the $(u,v)$
parametrisation remains unaltered. The
metric
\begin{equation}\label{6.1}
\sigma_{ik}=g_{ik}-\left(\frac{1}{\epsilon\mu}-1\right)u_{i}u_{k}
\end{equation}
takes in the Hamiltonian the role that $g_{ik}$ had in the
previous Section, but of course also the latter shall intervene:
it is the real metric of spacetime, and
{\it e.g.} the line elements $\dd s_0$ and $\dd s$ entering the four-velocities of
the atoms and of the medium must be measured by $g_{ik}$.
We define the null geodesics of the effective spacetime as
\begin{equation}\label{6.2}
\frac{^{\sigma}\delta}{\delta u}\frac{\dd x^i}{\dd u} \equiv
\frac{\dd ^2x^i}{\dd u^2}+\Sigma^i_{kl}\frac{\dd x^k}{\dd u}\frac{\dd x^l}{\dd u}=0,
\end{equation}
where
\begin{equation}\label{6.3}
\sigma_{ik}\frac{\dd x^i}{\dd u}\frac{\dd x^k}{\dd u}=0,
\end{equation}
and the Christoffel connection to be used is now
\begin{equation}\label{6.4}
\Sigma^i_{kl}=\frac{1}{2}\sigma^{im}\left(\sigma_{mk,l}+\sigma_{ml,k}
-\sigma_{kl,m}\right).
\end{equation}
Gordon's reduction to the vacuum imposes to model the photon
as a null particle of the effective spacetime, and to write
\begin{equation}\label{6.5}
M^i=\theta\frac{\dd x^i}{\dd u},\text{~~~}\frac{^{\sigma}\delta
M^i}{\delta u}=0,
\end{equation}
where $\theta$, assumed to be positive, happens to be a constant
thanks to (\ref{6.2}). Since the photon travels with the
fundamental velocity of $\sigma_{ik}$ we shall pose
\begin{equation}\label{6.6}
\sigma_{ik}M^iM^k\equiv M_{(i)}M^i=0,
\end{equation}
where we have adopted Gordon's convention of enclosing in round
parentheses the indices moved with the effective metric.
We can again model the interaction of the photon with the atom
through the equation $m'_0 {u'}^i=m_0 u^i+M^i$. Since
\begin{eqnarray}\label{6.7}
\sigma_{ik}M^iM^k=\left[g_{ik}+
\left(1-\frac{1}{\epsilon\mu}\right)u_iu_k\right]M^iM^k\\\nonumber
=g_{ik}M^iM^k+\left(1-\frac{1}{\epsilon\mu}\right)\left(u_iM^i\right)^2=0,
\end{eqnarray}
one finds
\begin{equation}\label{6.8}
g_{ik}M^iM^k\equiv
M_iM^i=\left(\frac{1}{\epsilon\mu}-1\right)\left(u_iM^i\right)^2.
\end{equation}
Since $\epsilon\mu>1$, when measured with the real spacetime metric,
the momentum-energy of the photon turns out to be timelike, as it occurs to
an ordinary particle, for which a rest coordinate system exists.
Therefore one finds
\begin{equation}\label{6.9}
-{m'}_0^2=-m_0^2+2m_0M_iu^i
+\left(\frac{1}{\epsilon\mu}-1\right)\left(u_iM^i\right)^2;
\end{equation}
instead of (\ref{5.5}), but for optical processes the last term is
negligible when compared to the remaining ones, and one can still define
the energy of the photon as in equation (\ref{5.7}). The argument
of the previous Section can now be repeated with the due
changes, by starting from  the quantity
\begin{equation}\label{6.10}
\frac{\partial}{\partial u}
\left(M_{(i)}\frac{\partial x^i}{\partial v}\right)
=\frac{^\sigma\delta M_{(i)}}
{\delta u}\frac{\partial x^i}{\partial v}
+M_{(i)}\frac{^\sigma\delta}
{\delta u}\frac{\partial x^i}{\partial v}.
\end{equation}
One obtains
\begin{equation}\label{6.11}
\frac{\partial}{\partial u} \left(M_{(i)}\frac{\partial x^i}
{\partial v}\right)=0,
\end{equation}
hence, again with reference to Figure 1:
\begin{equation}\label{6.12}
\left(M_{(i)}\frac{\partial x^i}{\partial v}\right)_{A_0}
=\left(M_{(i)}\frac{\partial x^i}{\partial v}\right)_A.
\end{equation}
By the same argument as in the previous Section:
\begin{equation}\label{6.13}
\left(M_{(i)}u^i\right)_{A_0}ds_0=\left(M_{(i)}u^i\right)_Ads,
\end{equation}
but, because
\begin{eqnarray}\label{6.14}
M_{(i)}u^i
=\left[g_{ik}+
\left(1-\frac{1}{\epsilon\mu}\right)u_iu_k\right]M^iu^k
=\frac{M_iu^i}{\epsilon\mu},
\end{eqnarray}
instead of (\ref{5.14}) one finds
\begin{equation}\label{6.15}
\left(\frac{M_iu^i}{\epsilon\mu}\right)_{A_0}ds_0
=\left(\frac{M_iu^i}{\epsilon\mu}\right)_Ads.
\end{equation}
Since equation (\ref{5.16}) remains unaltered, instead of (\ref{5.17})
one eventually obtains
\begin{equation}\label{6.16}
\left(\frac{E}{\nu\epsilon\mu}\right)_{A_0}
=\left(\frac{E}{\nu\epsilon\mu}\right)_A.
\end{equation}
\section{Spacelike photons in nondispersive, isotropic dielectrics}
The photon that we have found through the reduction to the vacuum
has the unquestionable virtue of being a timelike entity for which
a rest system exists and, as far as a homogeneous medium is
considered, it exhibits the proportionality of energy and
frequency that a good-natured photon is supposed to possess. But,
as soon as a nonhomogeneous medium is considered, the
proportionality disappears, and we are left wondering whether
we did some wrong, or whether there is some other opportunity
that we have missed.

    The Hamiltonian point of view tells us that the latter is the
case. In fact, with (\ref{4.8}) there is an
ambiguity in the choice of the momentum-energy vector, that is not
apparent in the vacuum case. Let us consider the first of
Hamilton's equations (\ref{2.6}). In our case it reads:
\begin{equation}\label{7.1}
\frac{\dd x^i}{\dd u}=\sigma^{ik}f_{,k}.
\end{equation}
Let
\begin{equation}\label{7.2}
p_k=\theta f_{,k}
\end{equation}
be the covariant version of the momentum-energy vector; it conforms
to the usual Hamiltonian prescription for material particles. Then
\begin{equation}\label{7.3}
M^i=\sigma^{ik}p_k.
\end{equation}
In the vacuum case, $M^i$ and $p_k$ are the contravariant and the
covariant forms of one and the same entity, whose norm is null.
When $\epsilon\mu>1$, $M^i$ and $p_i$ are expressions of different
entities: the first one is a timelike vector, while the second one
is spacelike.
Let us assume that not $M^i$, but $p_i$ is the momentum-energy of the
photon, by definition the one that is exchanged with an atom in the absorption
and emission processes.
Instead of (\ref{5.4}) we shall postulate
\begin{equation}\label{7.7}
m'_0u'_i=m_0u_i+p_i.
\end{equation}
If this equation is rewritten in contravariant form and the two
forms are multiplied term by term one obtains
\begin{equation}\label{7.8}
-{m'}_0^2=-m_0^2+2m_0p_iu^i+p_ip^i.
\end{equation}
For optical processes
one can neglect the last term and define
\begin{equation}\label{7.9}
\bar{E}=-p_iu^i
\end{equation}
as the energy exchanged by the photon. Due to (\ref{7.3}) one
writes
\begin{equation}\label{7.10}
M_iu^i=u_k\sigma^{kl}p_l
=u_k\left[g^{kl}-(\epsilon\mu-1)u^ku^l\right]p_l
=\epsilon\mu p_iu^i,
\end{equation}
hence, instead of the relation (\ref{6.16}), $\bar{E}$
fulfils
\begin{equation}\label{7.11}
\left(\frac{\bar E}{\nu}\right)_{A_0}
=\left(\frac{\bar E}{\nu}\right)_A.
\end{equation}
\section{Conclusion}
While general relativity in vacuo contains a satisfactory model of
the photon as a null particle whose world-line lies in a null wave, and for which
the proportionality between energy and frequency holds, the
situation is much less encouraging if one attempts defining, by general
relativistic methods, a photon in nondispersive, isotropic
dielectrics.

The Hamiltonian viewpoint shows that in such
dielectrics the photon should be a pointlike entity
associated to the ray of geometrical optics, that
in its turn is permanently associated to a given phase wave, but it
shows also that one must choose between two evils:
either maintaining the proportionality between energy and frequency
at the cost of associating a spacelike momentum to an entity that
travels along a timelike ray, or insisting on the natural association of a
timelike momentum to the timelike ray at the cost of abandoning,
when the dielectric is nonhomogeneous, the
proportionality of energy and frequency.
The old dilemma, born with the electromagnetic energy tensors of
Minkowski \cite{Minkowski1908} and Abraham \cite{Abraham1909}
and still awaiting for an experimental resolution \cite{Antoci1998},
thanks to the methods of Gordon \cite{Gordon1923}
and Synge \cite{Synge1935} appears here in a novel form.

\end{document}